\documentclass{eas}
\usepackage{graphicx}
\usepackage{natbib}
\usepackage{journals}
\begin{document}

\title{How do dwarf galaxies acquire their mass 
\\ \& when do they form their stars?} 
\runningtitle{Mass growth mode \& star formation history in dwarf galaxies} 
\author{G. A. Mamon}\address{Institut d'Astrophysique, Paris, France}
\author{D. Tweed}\sameaddress{1}\secondaddress{Institut d'Astrophysique
  Spatiale, Orsay, France}
\author{A. Cattaneo}\address{Astrophysikalisches Institut, Postdam, Germany}
\secondaddress{CRAL, Observatoire de Lyon, Lyon, France}
\author{T. X. Thuan}\address{Dept. of Astronomy, Univ. of Virginia,
  Charlottesville VA, USA}
\begin{abstract}
We apply a simple, one-equation, galaxy formation model on top of the halos
and subhalos of a high-resolution dark matter cosmological simulation to
study how dwarf galaxies acquire their mass and,
for better mass resolution, on over $10^5$ halo merger trees, to
predict  when they form their stars.
With the first approach, we show that the large majority of galaxies within
group- and cluster-mass halos have acquired the bulk of their stellar 
mass through gas
accretion and not via galaxy mergers. We deduce that most dwarf ellipticals are
not built up by galaxy mergers.
With the second approach, we constrain the star formation histories of dwarfs by
requiring that star formation must occur within halos of a minimum circular
velocity set by the evolution of the temperature of the IGM, starting before
the epoch of reionization.
We qualitatively reproduce
 the downsizing trend of greater ages at greater masses and
predict an upsizing trend of greater ages as one proceeds to masses lower
than $m_{\rm crit}$.
We find that the fraction of galaxies with very young stellar populations
(more than half the mass formed within the last 1.5 Gyr) is a
function of present-day mass in stars and cold gas, which peaks at 0.5\% at
$m_{\rm crit}$=$10^{6-8} M_\odot$, corresponding to blue compact dwarfs
such as I~Zw~18. 
We predict that the baryonic mass function of galaxies  
should not show a maximum at masses  above $10^{5.5}\,
M_\odot$,
and we
speculate on the nature of the lowest mass galaxies.
\end{abstract}
\maketitle
\section{Introduction}

There is still much debate on how galaxies acquire their mass and when do
they form their stars.
The mass growth of galaxies can occur 
either by 
 accretion of gas that cools to form molecular clouds in which stars form
or
by galaxy mergers.
While it is generally accepted that spiral disks form through the first mode,
it is still unclear whether elliptical galaxies are built by mergers or not.
We use a very simple \emph{toy} model of galaxy formation \citep{CMWK10} run on top of dark
matter halo merger trees to understand how dwarf galaxies
acquire their mass. We also use the toy model to predict the frequency of galaxies
with stellar population as young as the very metal-poor (1/50th solar
metallicity) galaxy I~Zw~18, for which analyses of the color-magnitude
diagrams observed with HST have revealed that 
the bulk of the stellar population is younger than 500 Myr \citep{IT04} or 1.5
Gyr (\citealp{TYI10}, see also \citealp{Aloisi+07}).

\section{How do galaxies acquire their mass?}

\begin{figure}
\centering
\includegraphics[width=7cm]{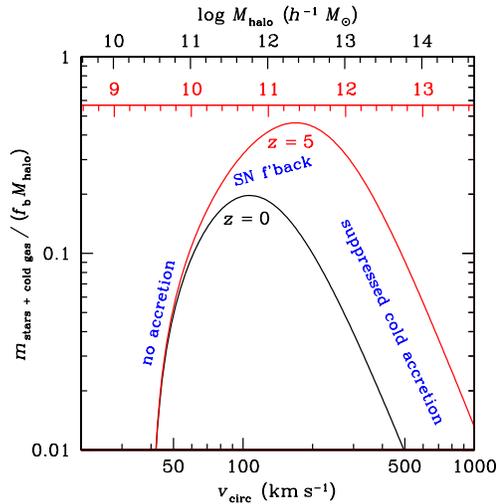}
\caption{Illustration \citep{CMWK10} of the toy model of galaxy formation
  (eq.~[\ref{toy}] with $v_{\rm reion}$=$40 \, \rm km \, s^{-1}$, $v_{\rm
    SN}$=$120 \, \rm km \, s^{-1}$, $M_{\rm shock}$=$8\times
  10^{11}\,h^{-1} M_\odot$ 
  and 40\% of the stellar mass stripped at every passage through a parent
  halo)  at two different 
  epochs: $z$=$0$ (upper halo mass scale) and $z$=$5$ (lower halo mass
  scale). The variations with redshift come from the redshift modulation of
  the dependence of $v_{\rm circ}$ on $M$.
\label{sfe}}
\end{figure}

Galaxies form in DM halos, and 
our
toy model 
gives the mass in stars and cold gas, $m$, as a function of halo
mass $M$ and epoch $z$, taking into account the fact that for stars to form
one needs: 1) gas accretion, which is fully quenched for low-mass halos
\citep{TW96,Gnedin00}; 2) in cold form, which becomes inefficient in
high-mass halos \citep{BD03,Keres+09a}; and 3) to retain the interstellar
gas against supernova (SN) winds \citep{DS86}:
\begin{equation}
m_{\rm stars\ +\ cold\ gas}(M,z)={v_{\rm circ}^2-v_{\rm reion}^2 \over v_{\rm circ}^2 + v_{\rm
    SN}^2}\,{f_{\rm b}\,M \over 1 + M/M_{\rm shock}} \ ,
\label{toy}
\end{equation}
where $f_{\rm b}$=$\Omega_{\rm b}/\Omega_{\rm m}\simeq 0.17$ 
is the cosmic baryon fraction,
$v_{\rm reion}$ is
the minimum halo circular velocity for gas accretion,
$v_{\rm SN}$ is a characteristic velocity for
SN feedback, 
and 
$M_{\rm shock}$ represents the transition from 
pure
cold 
to mainly hot accretion.
%
Figure~\ref{sfe} describes the efficiency of galaxy formation, $m_{\rm
  stars\ +\ cold\ gas}/(f_{\rm b} M)$, at $z$=0 and 5, with the parameters tuned to 
  match the observed ($z$$\simeq$0.1)
galaxy 
stellar mass function (MF) of \cite{BMcIKW03}. 
Galaxy formation 
occurs in a fairly narrow range of halo masses, that varies with redshift.

We apply equation~(\ref{toy}) to the merger trees run on the halos and
subhalos (AHF algorithm of \citealp{KK09}) 
of a high resolution dark matter cosmological simulation (CS)
When a halo enters a
more massive one, it becomes a \emph{subhalo} and its galaxy becomes a
\emph{satellite}. The orbit of the subhalo is followed 
until either 1) dynamical friction (DF) causes it to fall to the halo
center and necessarily see its galaxy merge with the central galaxy of the
parent halo, or 2) it is tidally stripped and heated by the global potential of the halo
to the point that there are insufficient particles to follow it. In this
latter case ($M_{\rm subhalo} < 1.5$$\times$$10^9 \,h^{-1} M_\odot$,
corresponding to $v_{\rm circ}=17 \, \rm km \, s^{-1}$ at $z=0$), 
we assume that the satellite galaxy merges with the central one
after a delay set by DF, for which we adopt the
timescale carefully 
calibrated by \cite{Jiang+08} with hydrodynamical CSs. 

\begin{figure}
\centering
\includegraphics[angle=-90,width=7cm]{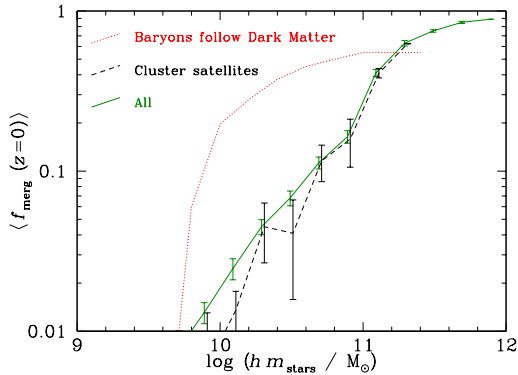}
\caption{Median fraction of $z$=0 stellar mass (for $h m_{\rm stars} >
  10^{10} M_\odot$ one can neglect the gas) acquired by mergers, for all galaxies
  (\emph{solid green}) and $h\,M_{\rm halo}> 10^{13} M_\odot$ cluster
  satellites (\emph{dashed black}). The \emph{dotted red line} shows the
  baryons trace the dark matter model.
The error bars are
  uncertainties on the median from 100 bootstraps. 
\label{fmvsm}}
\end{figure}
Figure~\ref{fmvsm} 
shows that
while mergers dominate the growth of the massive galaxies (as expected from
the toy model, since gas accretion is quenched at high masses), their
importance drops sharply when one moves to masses below $10^{11}\,h^{-1}
M_\odot$ (the mass resolution is estimated at $10^{10.6} \,h^{-1} M_\odot$,
where the median fraction of mass acquired by mergers no longer decreases
with mass faster than in our reference model where baryons trace the dark
matter [red dotted line]).
This dominance of gas accretion at low mass is also true for the satellites
of clusters (dashed line). Since observations indicate that most satellites
of clusters are 
dwarf ellipticals (dEs), we conclude that cluster dEs are most often 
not built by mergers. One must resort to other mechanisms (not included in
our toy model) that transform dwarf irregulars into dEs
(harassment [\citealp{Mastropietro+05}] or ram pressure stripping [\citealp{BBCG08}]).

\section{When do dwarf galaxies form their stars?}

\begin{figure}
\centering
\includegraphics[angle=-90,width=0.85\hsize,bb=120 1 506 760]{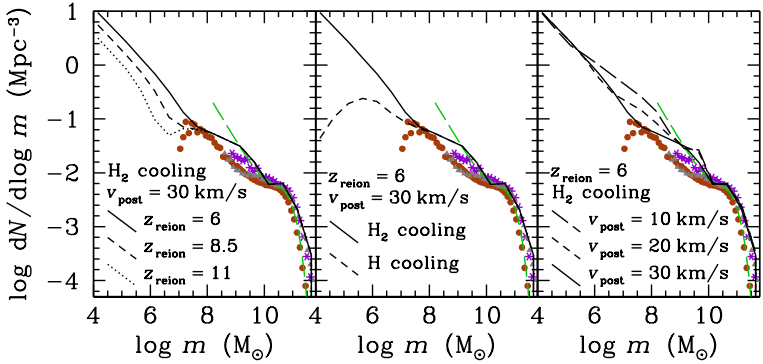}
\caption{Effects of thermal history of IGM on galaxy \emph{baryonic} mass function.
\emph{Left:}
Effect of reionization epoch.
\emph{Middle:}
Effect of pre-reionization IGM temperature.
\emph{Right:}
Effect of post-reionization IGM temperature.
In all plots, the \emph{green long-dashed lines} and \emph{symbols} respectively 
represent the observed SDSS \emph{baryonic} \citep{BGD08} and \emph{stellar} mass
functions, 
measured by 
\cite{BMcIKW03} (\emph{purple asterisks}),
\cite{BGD08} (\emph{brown circles}), 
and \cite{YMvdB09} (\emph{gray triangles}).
\label{masfuns}}
\end{figure}

In our toy model of galaxy formation, galaxy formation can only occur in
halos above a critical mass (corresponding to $v_{\rm circ}>v_{\rm reion}$).
We have used the halo merger tree code of \cite{ND06} to statistically study
the star formation histories (SFHs) of dwarf galaxies. We consider 
24 final halo masses geometrically spaced between $10^7$ and $10^{12.75}
\,h^{-1} M_\odot$, and run each halo merger tree 10$\,$000 times.
We run the toy model on top of the branches of the halo merger tree (moving
forward in time) to follow the evolution of stellar mass (plus cold gas).

%
The predicted \emph{baryonic} (stars + cold gas) MFs generally do not match
that observed in the SDSS \citep{BGD08},
(but by coincidence they do
match the observed \emph{stellar} MFs).
The left panel of Figure~\ref{masfuns}
suggests that reionization must occur late ($z$=6).
The middle panel of Figure~\ref{masfuns} 
hints that, before reionization, the temperature of the IGM
must be set by molecular cooling ($v_{\rm pre-reion}$=$2 \, \rm km \, s^{-1}$).
The right panel indicates a good match between predicted and observed
baryonic MFs 
when the IGM after
reionization is fairly cool ($v_{\rm post-reion} = 10 \, \rm
km \, s^{-1}$).

\begin{figure}
\includegraphics[width=0.49\hsize]{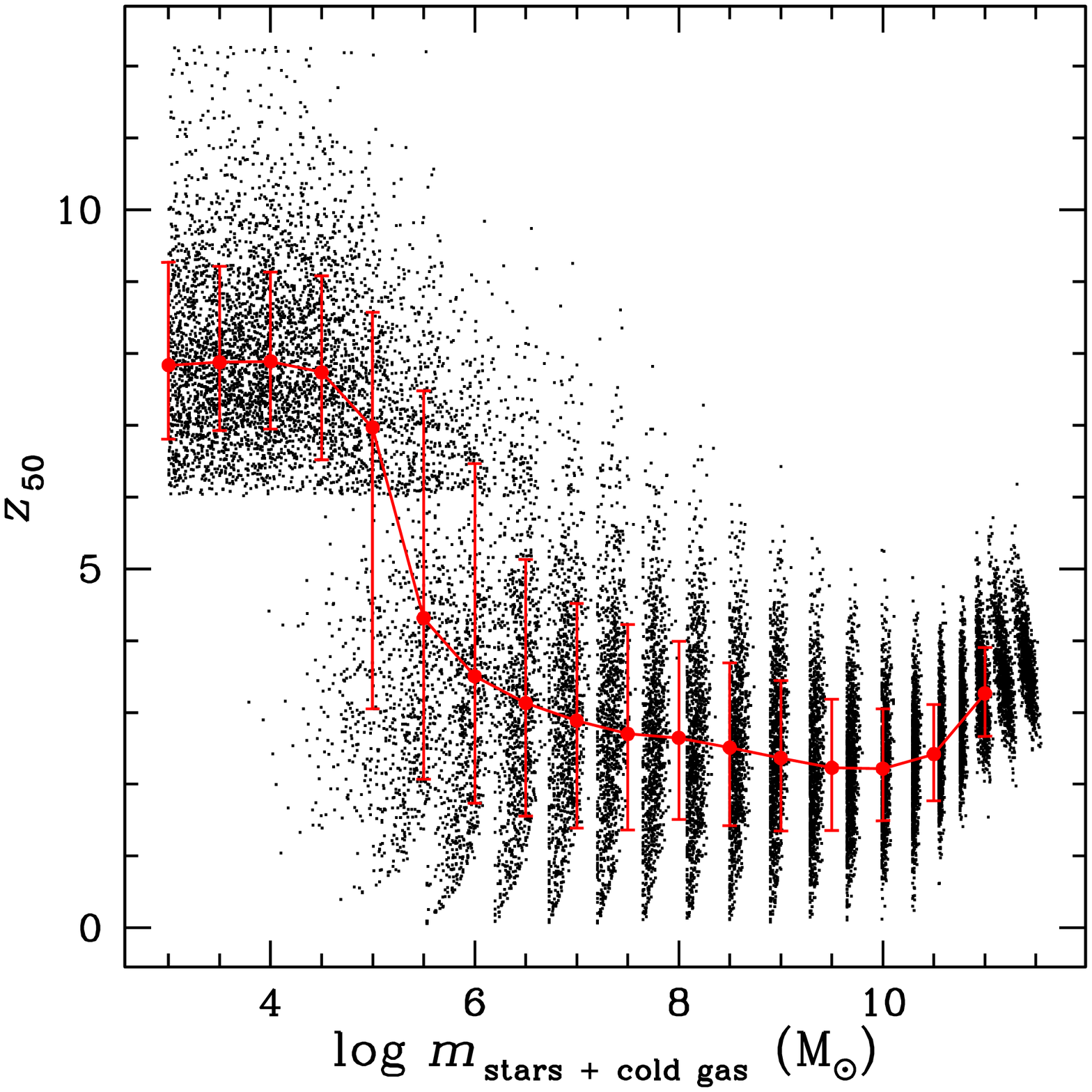}
\quad
\includegraphics[width=0.49\hsize]{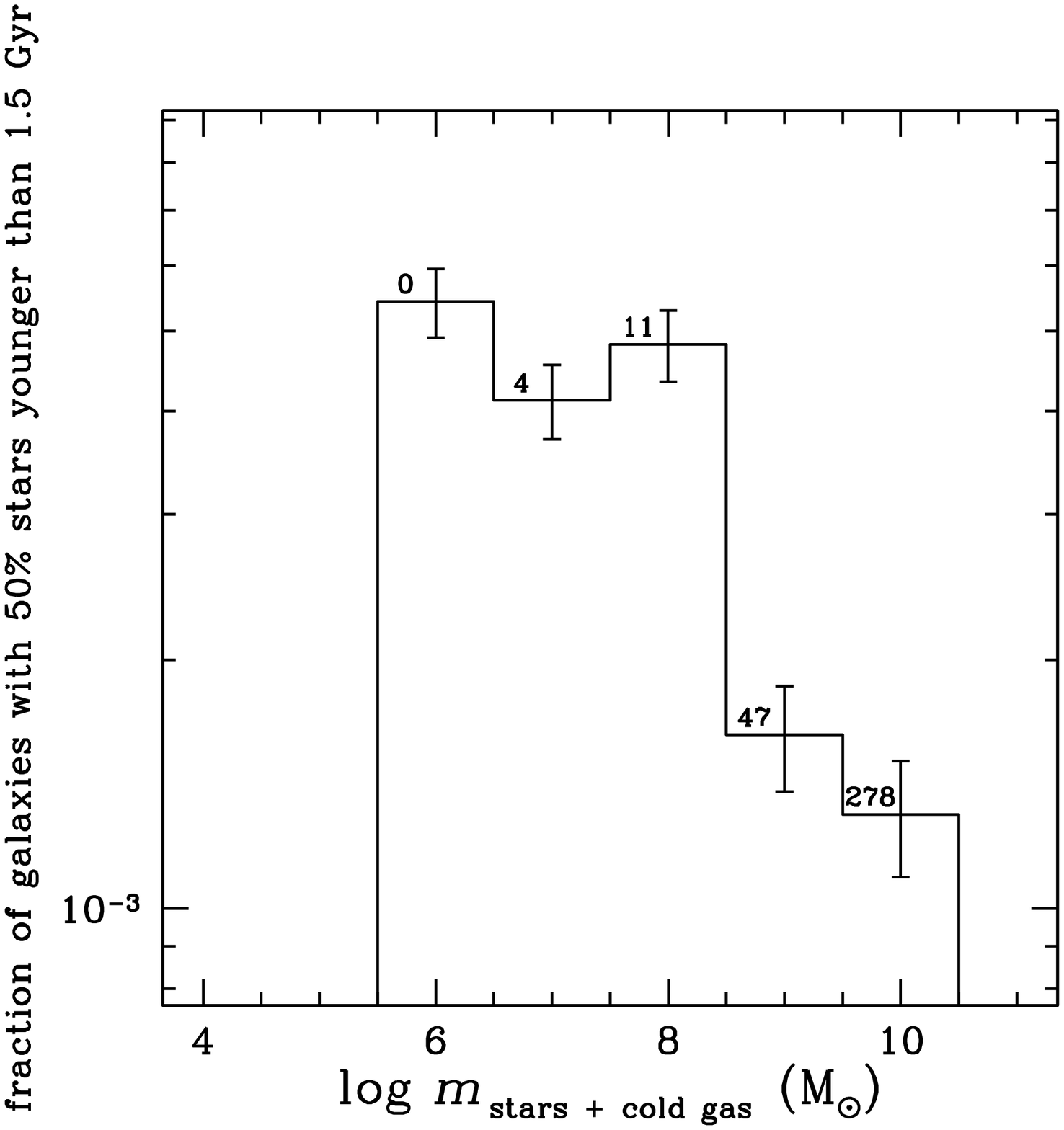}
\caption{\emph{Left:} 
Median (mass-weighted) star formation redshift vs $z$=0 mass. The points (1
in 5 plotted for clarity) are
individual galaxies, while the red symbols are medians (error bars extend from
16th to 84th percentiles). The stripes are artifacts
of our discrete set of final halo masses.
\emph{Right:}
Fraction of young galaxies (half the mass formed more recently than
  last 1.5 Gyr) versus $z$=0 baryonic mass (with
numbers of expected SDSS galaxies).
In both plots, we adopt $v_{\rm pre-reion}$=$2 \, \rm km \, s^{-1}$ ($\rm H_2$
cooling) before
reionization ($z=6$), $v_{\rm post-reion}=10 \, \rm km \, s^{-1}$ for $z<6$,
$v_{\rm SN}=120 \, \rm km \, s^{-1}$, and
$h M_{\rm shock}=8\!\times\!10^{11} \,M_\odot$. 
}
\label{fracyoung}
\end{figure}

According to
the left panel of Figure~\ref{fracyoung},
as one
proceeds from the highest final stellar (plus cold gas) masses to lower ones,
one first notices that the median stellar age diminishes, qualitatively
reproducing the \emph{downsizing} of star formation. However this downsizing
stops at masses of $10^{8-10}\,h^{-1} M_\odot$ (roughly the masses of blue
compact dwarfs) and as one proceeds to even
lower masses, one notices an \emph{upsizing} of stellar ages. In our model, 
the smallest galaxies
have the bulk of their stars formed before reionization.

The right panel of Figure~\ref{fracyoung} shows that the frequency of
galaxies with the bulk of their mass in stars (and cold gas remaining in the
galaxy) acquired within the last 1.5 Gyr is maximal at  
$\simeq 0.5\%$ at $m_{\rm stars+cold\,gas}$=$10^{6-8}
M_\odot$.  The presence of a young galaxy such as I~Zw~18, whose
baryonic mass is of order $10^{7.9} M_\odot$ \citep{TYI10}, is consistent
with our model (and robust to the parameters).
However, our model predicts as many as 340 young galaxies in the SDSS, mostly at
$m_{\rm stars\ +\ cold\ gas} = 10^{10} M_\odot$ (and twice as many with
$v_{\rm post-reion} = 30 \, \rm km \, s^{-1}$), which may present a
challenge for such modeling.

\section{The lowest mass galaxies}

The middle panel of Figure~\ref{masfuns} indicates that there is no peak in
the best fitting galaxy (stars plus cold gas) MF, if the IGM
temperature before reionization is set by molecular
Hydrogen cooling ($v_{\rm pre-reion}$=$2 \, \rm km \, s^{-1}$) with a low-end
slope of $-1.55$ (in comparison the baryonic MF measured by \citealp{BGD08}
has a slope of $-1.87$). If, instead, the IGM temperature before reionization is set
by atomic Hydrogen cooling ($v_{\rm pre-reion}$=$17 \, \rm km \, s^{-1}$), the mass
function peaks at $m_{\rm stars+cold\,gas}$=$10^{5.5} \,h^{-1}
M_\odot$. This maximum is probably not caused by of our mass resolution,
since no such peak is seen when the pre-reionization IGM temperature is set by
$\rm H_2$ cooling.

The importance of the low-end tail of the galaxy MF  raises the
question of the nature of very low mass objects ($m_{\rm stars + cold\,gas} <
10^6 M_\odot$).
Two classes of objects come to mind: Globular Clusters (GCs) and High Velocity
Clouds (HVCs). However, in our model, these objects must be (or have been)
associated with DM halos. While Galactic HVCs do appear to require DM
\citep{BW04},
Galactic GCs don't
(e.g. \citealp{Sollima+09}), perhaps because they are closer and more tidally stripped.


\small

\bibliography{master}

\begin{thebibliography}{19}
\expandafter\ifx\csname natexlab\endcsname\relax\def\natexlab#1{#1}\fi

\bibitem[{{Aloisi} {et~al.}(2007){Aloisi}, {Clementini}, {Tosi}, {Annibali},
  {Contreras}, {Fiorentino}, {Mack}, {Marconi}, {Musella}, {Saha}, {Sirianni},
  \& {van der Marel}}]{Aloisi+07}
{Aloisi}, A., {Clementini}, G., {Tosi}, M., {et~al.} 2007, \apjl, 667, L151

\bibitem[{{Baldry} {et~al.}(2008){Baldry}, {Glazebrook}, \& {Driver}}]{BGD08}
{Baldry}, I.~K., {Glazebrook}, K., \& {Driver}, S.~P. 2008, \mnras, 388, 945

\bibitem[{{Bell} {et~al.}(2003){Bell}, {McIntosh}, {Katz}, \&
  {Weinberg}}]{BMcIKW03}
{Bell}, E.~F., {McIntosh}, D.~H., {Katz}, N., \& {Weinberg}, M.~D. 2003, \apjs,
  149, 289

\bibitem[{{Birnboim} \& {Dekel}(2003)}]{BD03}
{Birnboim}, Y. \& {Dekel}, A. 2003, \mnras, 345, 349

\bibitem[{{Boselli} {et~al.}(2008){Boselli}, {Boissier}, {Cortese}, \&
  {Gavazzi}}]{BBCG08}
{Boselli}, A., {Boissier}, S., {Cortese}, L., \& {Gavazzi}, G. 2008, \apj, 674,
  742

\bibitem[{{Br{\"u}ns} \& {Westmeier}(2004)}]{BW04}
{Br{\"u}ns}, C. \& {Westmeier}, T. 2004, \aap, 426, L9

\bibitem[{{Cattaneo} {et~al.}(2010){Cattaneo}, {Mamon}, {Warnick}, \&
  {Knebe}}]{CMWK10}
{Cattaneo}, A., {Mamon}, G.~A., {Warnick}, K., \& {Knebe}, A. 2010, \aap,
  submitted, arXiv:1002.3257

\bibitem[{{Dekel} \& {Silk}(1986)}]{DS86}
{Dekel}, A. \& {Silk}, J. 1986, \apj, 303, 39

\bibitem[{{Gnedin}(2000)}]{Gnedin00}
{Gnedin}, N.~Y. 2000, \apj, 542, 535

\bibitem[{{Izotov} \& {Thuan}(2004)}]{IT04}
{Izotov}, Y.~I. \& {Thuan}, T.~X. 2004, \apj, 616, 768

\bibitem[{{Jiang} {et~al.}(2008){Jiang}, {Jing}, {Faltenbacher}, {Lin}, \&
  {Li}}]{Jiang+08}
{Jiang}, C.~Y., {Jing}, Y.~P., {Faltenbacher}, A., {Lin}, W.~P., \& {Li}, C.
  2008, \apj, 675, 1095

\bibitem[{{Kere{\v s}} {et~al.}(2009){Kere{\v s}}, {Katz}, {Fardal},
  {Dav{\'e}}, \& {Weinberg}}]{Keres+09a}
{Kere{\v s}}, D., {Katz}, N., {Fardal}, M., {Dav{\'e}}, R., \& {Weinberg},
  D.~H. 2009, \mnras, 395, 160

\bibitem[{{Knollmann} \& {Knebe}(2009)}]{KK09}
{Knollmann}, S.~R. \& {Knebe}, A. 2009, \apjs, 182, 608

\bibitem[{{Mastropietro} {et~al.}(2005){Mastropietro}, {Moore}, {Mayer},
  {Debattista}, {Piffaretti}, \& {Stadel}}]{Mastropietro+05}
{Mastropietro}, C., {Moore}, B., {Mayer}, L., {et~al.} 2005, \mnras, 364, 607

\bibitem[{{Neistein} \& {Dekel}(2008)}]{ND06}
{Neistein}, E. \& {Dekel}, A. 2008, \mnras, 383, 615

\bibitem[{{Sollima} {et~al.}(2009){Sollima}, {Bellazzini}, {Smart}, {Correnti},
  {Pancino}, {Ferraro}, \& {Romano}}]{Sollima+09}
{Sollima}, A., {Bellazzini}, M., {Smart}, R.~L., {et~al.} 2009, \mnras, 396,
  2183

\bibitem[{{Thoul} \& {Weinberg}(1996)}]{TW96}
{Thoul}, A.~A. \& {Weinberg}, D.~H. 1996, \apj, 465, 608

\bibitem[{{Thuan} {et~al.}(2010){Thuan}, {Yakobchuk}, \& {Izotov}}]{TYI10}
{Thuan}, T.~X., {Yakobchuk}, T.~M., \& {Izotov}, Y.~I. 2010, \apj, submitted

\bibitem[{{Yang} {et~al.}(2009){Yang}, {Mo}, \& {van den Bosch}}]{YMvdB09}
{Yang}, X., {Mo}, H.~J., \& {van den Bosch}, F.~C. 2009, \apj, 695, 900

\end{thebibliography}
\end{document}